\documentclass[prl,twocolumn,superscriptaddress,showpacs,preprintnumbers]{revtex4}
\usepackage{bm,graphicx}

\begin{document}

\title{Geometric depolarization in patterns formed by backscattered light}

\author{David Lacoste}
\affiliation{Laboratoire de Physico-Chimie Th\'eorique, ESPCI, 10
rue Vauquelin, F-75231 Paris Cedex 05, France}
\author{Vincent Rossetto}
\affiliation{Laboratoire de Physique et Mod\'elisation des Milieux
Condens\'es, Maison des Magist\`eres, B.P. 166, 38042 Grenoble
Cedex 9, France}
\author{Franck Jaillon}
\affiliation{Laboratoire de R\'esonance Magn\'etique Nucl\'eaire,
M\'ethodologie et Instrumentation en Biophysique, UMR CNRS 5012,
Domaine de la Doua - CPE - 3, rue Victor Grignard, 69616
Villeurbanne, France}
\author{Herv\'e Saint-Jalmes}
\affiliation{Laboratoire de R\'esonance Magn\'etique Nucl\'eaire,
M\'ethodologie et Instrumentation en Biophysique, UMR CNRS 5012,
Domaine de la Doua - CPE - 3, rue Victor Grignard, 69616
Villeurbanne, France}

\begin{abstract}
We formulate a framework for the depolarization of linearly
polarized backscattered light based on the concept of geometric
phase, {\it i.e} Berry's phase. The predictions of this theory are
applied to the patterns formed by backscattered light between
crossed or parallel polarizers. This theory should be particularly
adapted to the situation in which polarized light is scattered
many times but predominantly in the forward direction. We apply
these ideas to the patterns which we obtained experimentally with
backscattered polarized light from a colloidal suspension.
\end{abstract}

\maketitle

\newpage
The transport of light through human tissues is one of the most
promising technique to detect in a noninvasive way for instance
breast cancer. For medical imaging applications, it is important
to extract the information contained not only in the intensity but
also in the polarization of backscattered light. This is not easy
in general due to the complexity of vector-wave multiple
scattering. In this paper we study a simple experiment, in which
polarized light is backscattered from a diffuse medium. In these
conditions, one observes between crossed polarizers a fourfold
symmetry pattern which was first interpreted qualitatively by
Dogariu and Asakura \cite{dogariu2}. Recently more quantitative
approaches have been developed for Mie scatterers using Mueller
matrices \cite{hielscher2}. A rather good agreement has been found
between the experimental shapes of the patterns and the
theoretically predicted ones \cite{rakovic,hielscher2,jaillon}.

In this paper, we propose an alternate approach, which is very
simple to implement because it is not based on a vector radiative
transfer method as used generally in the literature. Instead our
approach is based on the notion of geometric phase, which was
introduced by Berry \cite{Berry} in his interpretation of the
experiments showing optical activity in an helically wound optical
fiber \cite{tomita}. So far, the concept has been mostly applied
to quantum mechanics and to field theory but has not been used in
the context of the transport of polarization in random media
except in the recent ref.~\cite{Maggs}, which we follow and extend
in this paper. Before presenting our framework, we discuss the
cross-shaped patterns using the Stokes formalism to make contact
with previous work \cite{rakovic,hielscher2,jaillon}.

The Stokes parameters are the elements of the vector ${\bf
I}=\left( I,Q,U,V \right)$ which is defined with respect to a
plane of reference containing the direction of propagation. If the
plane of reference is rotated through an angle $\phi$, the new
Stokes vector is $L(\phi) {\bf I}$ where $L(\phi)$ is a rotation
matrix \cite{rakovic}. We used the scattering matrix $S$
corresponding to a distribution of spherical scatterers. The
incident light is linearly polarized, is described by a Stokes
vector $\bf I_0$ and is normal to the medium. Since the medium has
cylindrical symmetry, the $S$ matrix is independent of $\phi$.
After going through an analyzer (with Mueller matrix $A$), the
outgoing Stokes vector is $A \cdot L(\phi) \cdot S \cdot L(\phi)
\cdot {\bf I}_0$. This means that the outgoing intensity in the
crossed polarized (resp. copolarized) channel is
\begin{equation}\label{I_r}
I_\perp=\frac{1}{4} \left( 2 S_{11} + S_{33} - S_{22} \right)
-\frac{1}{4} \left( S_{22}+S_{33} \right) \cos 4\phi,
\end{equation}
\begin{equation}\label{I_l}
I_\parallel=\frac{1}{4} \left( 2 S_{11} - S_{33} + S_{22} \right)
- S_{12} \cos 2 \phi+\frac{1}{4} \left( S_{22}+S_{33} \right) \cos
4\phi,
\end{equation}
corresponding to Stokes parameters $I=I_\perp+I_\parallel$ and
$Q=I_\parallel-I_\perp$. Note that according to Eq.~\ref{I_r} the
crossed-polarized pattern has a four-fold symmetry, whereas
according to Eq.~\ref{I_l}, we see that the copolarized pattern
contains in addition a two-fold symmetry due to the term
proportional to $S_{12}$ \cite{jaillon}. In the particular case,
which is satisfied in multiple light scattering \cite{freund},
when $S=S_{11} diag(1,C,C,D)$, with $C=S_{22}/S_{11}$ and
$D=S_{44}/S_{11}$, Eqs.~\ref{I_r}-\ref{I_l} take the simple form
\begin{equation}\label{I_rsimple}
I_\perp=\frac{1}{2} I_{0} \left(1 - C \cos 4 \phi \right),
\end{equation}
\begin{equation}\label{I_lsimple}
I_\parallel=\frac{1}{2} I_{0} \left(1 + C \cos 4 \phi \right),
\end{equation}
corresponding to outgoing Stokes parameters $I=I_0=S_{11}$ and
$Q=C I_0 \cos 4 \phi$. Note that a cross is expected now in both
polarization channels and that $C$ measures the contrast of these
patterns. We have shown here that the patterns follow from general
properties of symmetry independently of the order of scattering or
of the degree of coherence of the source.

Let us now discuss the geometric depolarization of linearly
polarized light. For Rayleigh scattering, the (linear)
polarization vector after scattering $\bf E'$ is $\bf E'=\bf k'
\times \left( \bf E \times \bf k' \right)$, in terms of the
polarization vector before scattering $\bf E$ and the scattered
wavevector $\bf k'$. This implies that $\bf E$ evolves by parallel
transport in the limit of small scattering angles, and diffuse on
the sphere of wavevector directions until the memory of the
polarization has been lost. Akkermans et al. has shown that this
leads to a depolarization with a characteristic depolarization
length $\ell_p$ equal to $\ell/\ln (10/7) \simeq 2.8 \ell$
\cite{akkermans}. Recently $\ell_p$ was measured using
polarization resolved DWS \cite{luis}, which confirmed Akkerman's
prediction for Rayleigh scatterers, and which gave $\ell_p \simeq
\ell^*$ in the limit of forward-peaked scattering $g \rightarrow
1$ in agreement with Monte Carlo simulations \cite{brosseau}. Here
we assume forward-peaked scattering because it applies to many
biological tissues, and because in this case there is a clear
analogy between light scattering paths and semi-flexible polymers
\cite{Maggs}. Recently, this analogy has been put on a solid
basis, by realizing that the Fokker-Planck (FP) equation which
describes semi-flexible polymers can be derived from the radiative
transfer equation in this limit \cite{Keller}. In the following,
we carry further the analogy by discussing the degrees of freedom
of twist (polymer) analogous to polarization (light scattering).

Let us consider a path of light, which we assume to be normally
incident on a semi-infinite random medium. Following
ref.~\cite{Berry}, we express the polarization vector $\bf E$ in a
basis of two vectors $({\bf n},{\bf b})$ normal to the tangent
vector $\bf u$ (if the path is regular enough, the Fr\'enet frame
is a possible choice) as shown in figure 1:
\begin{equation}\label{Frenet}
{\bf E}(t)=c_1(t) {\bf n}(t) + c_2(t) {\bf b}(t),
\end{equation}
where $t$ is a parameter which goes from 0 to $s$ along the path.
Let us call $\phi$, the angle between ${\bf E}$ and ${\bf n}$ at
$t=0$, so that $c_2(0)/c_1(0)=\tan \phi$. Since the polarization
evolves by parallel transport $\dot{c_1}=\tau c_2$ and
$\dot{c_2}=-\tau c_1$, where $\tau$ denotes the torsion on the
trajectory, as found many years ago by Rytov \cite{Rytov}. In the
backscattering geometry, ${\bf n}(t=s)=-{\bf n}(t=0)$ and ${\bf
b}(t=s)={\bf b}(t=0)$, therefore we find that the polarization
vector at the end of the path is
\begin{equation}\label{E_fin}
{\bf E}(t=s)=-\cos (\phi+\Omega(s)) {\bf n}(s) + \sin
(\phi+\Omega(s)) {\bf b}(s),
\end{equation}
where $\Omega(s)$ is a geometrical phase, equal to the opposite of
the integral of the torsion between $t=0$ and $t=s$ modulo $4\pi$
\cite{Berry}. In the analogy between a path of light and a
semi-flexible polymer, the twist of the path is zero for light (it
would be non-zero only in chiral medium), and the writhing angle
is precisely $\Omega$. This writhe is a real value since the path
is open, and that value is equal to the algebraic area of a random
walk on a unit sphere, with the constrain that the path goes from
the north pole to the south pole in the backscattering geometry.
From Eq.~\ref{E_fin}, we find that the output intensity after the
light has gone through an analyzer crossed with respect to the
direction of the incident polarization, is proportional to $\sin^2
(2\phi + \Omega)$. Because the medium is random, this intensity
must be averaged with respect to all paths :
\begin{equation}\label{I_integ}
I_\perp \left( R \right)=\int P'(s,R)ds < \sin^2 (2\phi +
\Omega(s))>,
\end{equation}
where $P'(s,R)$ is the distribution of path length for a given
distance to the incident beam $R$, and $<..>$ denotes the average
over paths of length $s$. After expanding the r.h.s of
Eq.~\ref{I_integ}, we obtain the form of Eq.~\ref{I_rsimple} with
$I_0(R)=\int P'(s,R)ds$ and the contrast is
\begin{equation}\label{CofR}
C(R) = \frac{1}{I_0(R)} \int  P'(s,R)ds < \cos \left( 2 \Omega(s) \right) >.\\
\end{equation}
The factor $\cos(2\Omega)$ in Eq.~\ref{CofR} means that the
contrast results from grouping pairs of paths of opposite
geometrical phases, and the sum over $s$ means that the phases of
any other paths are uncorrelated. Interestingly a similar
dephasing occurs in the theory of magneto-conductance of Anderson
insulators \cite{bouchaud}.

To evaluate the distributions of $\Omega$ for fixed $s$,
$P(s,\Omega)$ shown in fig.~2, we use a Monte Carlo algorithm
originally developed for semi-flexible polymers. Random paths are
generated with an exponential distribution of path length with a
characteristic step equal to the elastic mean free path $\ell$.
The incident photons are normal to the interface, but when exiting
the medium all outgoing angles of emergent photons are accepted.
The paths can be generated for an arbitrary ratio of the transport
mean free path to the elastic mean free path $\ell^*/\ell$. The
geometric phase are calculated by closing the paths on the
momentum sphere with a geodesic \cite{Maggs}. Because of this
closure, the distribution of $\Omega$ for short paths $s \ll
\ell^*$ is peaked at zero as is also found for planar random walks
(Levy's law). For long path $s \gg \ell^*$, the distribution of
$\Omega$ widens, until the polarization is completely lost. In
this regime, the distribution $P(s,\Omega)$ is gaussian as
required by the central limit theorem (the variance which was
quadratic for $s \ll \ell^*$ becomes linear for $s \gg \ell^*$ as
seen in fig.~2), which implies that $<\cos 2 \Omega(s) >$ is a
decreasing exponential function of $s$. In Fig.~3, we show the
corresponding curve for the contrast of the pattern calculated
from Eq.~\ref{CofR}, together with experimental points, which we
obtained by averaging the Stokes parameter $Q$ of an image along
two perpendicular directions thereby suppressing a possible
contribution in $\cos(2\phi)$ present in Eq.~\ref{I_l}. In the
experiment, a colloidal suspension of latex particles of
negligible absorption (diameter 0.5$\mu$m, wavelength
$\lambda=670$nm) was used and the sample was about $8.8 \ell^*$
thick \cite{jaillon}. The value of the anisotropy parameter $g$ in
the simulation was chosen to match the experimental value $g
\simeq 0.82$. In this figure, we see that the contrast decreases
exponentially as function of the distance $R$ with a
characteristic distance of the order of $\ell_p \simeq \ell^*$
both in the theory and in the experiments in agreement with
refs.~\cite{luis}-\cite{brosseau}. In the central region of the
pattern, low order scattering is dominant as confirmed
numerically. This could explain the discrepancy between
experiments and simulations in this region, since our model only
treats low order scattering events in an approximative way.

To conclude, we have developed in this paper a simple theoretical
framework for geometric depolarization, which we have applied to
the experimental backscattering patterns. The mechanism of
geometric depolarization is general provided that the scattering
is peaked in the forward direction. We hope that our work will
motivate further theoretical studies on the role of geometric
phases in the transport properties of polarization in random media. \\

We acknowledge many stimulating discussions with T. Maggs, M.
Cloitre, F. Monti, and B. A. van Tiggelen.


\newpage

\section*{List of Figures}
Fig.~1: Representation of a typical path in a semi-infinite random
medium in backscattering. The Fr\'enet frame consists of the
tangent $\bf u$, the normal $\bf n$, and the binormal $\bf b$
vectors. $R$ denotes the distance between end points, $\phi$ is
the initial angle between the polarization vector $\bf E$ and the
normal $\bf n$, and $\Omega$ is the geometric phase. \\

Fig.~2:  Distribution of the geometric phase $\Omega$ for
different values of the path length $s$ and in the inset variance
of the distribution as function of $s/\ell^*$. \\

Fig.~3: Contrast as function of $R$: the crosses have been
obtained from Monte-Carlo simulations using Eq.~\ref{CofR} and the
squares are experimental values, obtained from an analysis of the
Stokes parameter $Q$. \\

\newpage
\begin{figure} \label{fig:schema}
{\par\centering
{\rotatebox{0}{\includegraphics[scale=0.5]{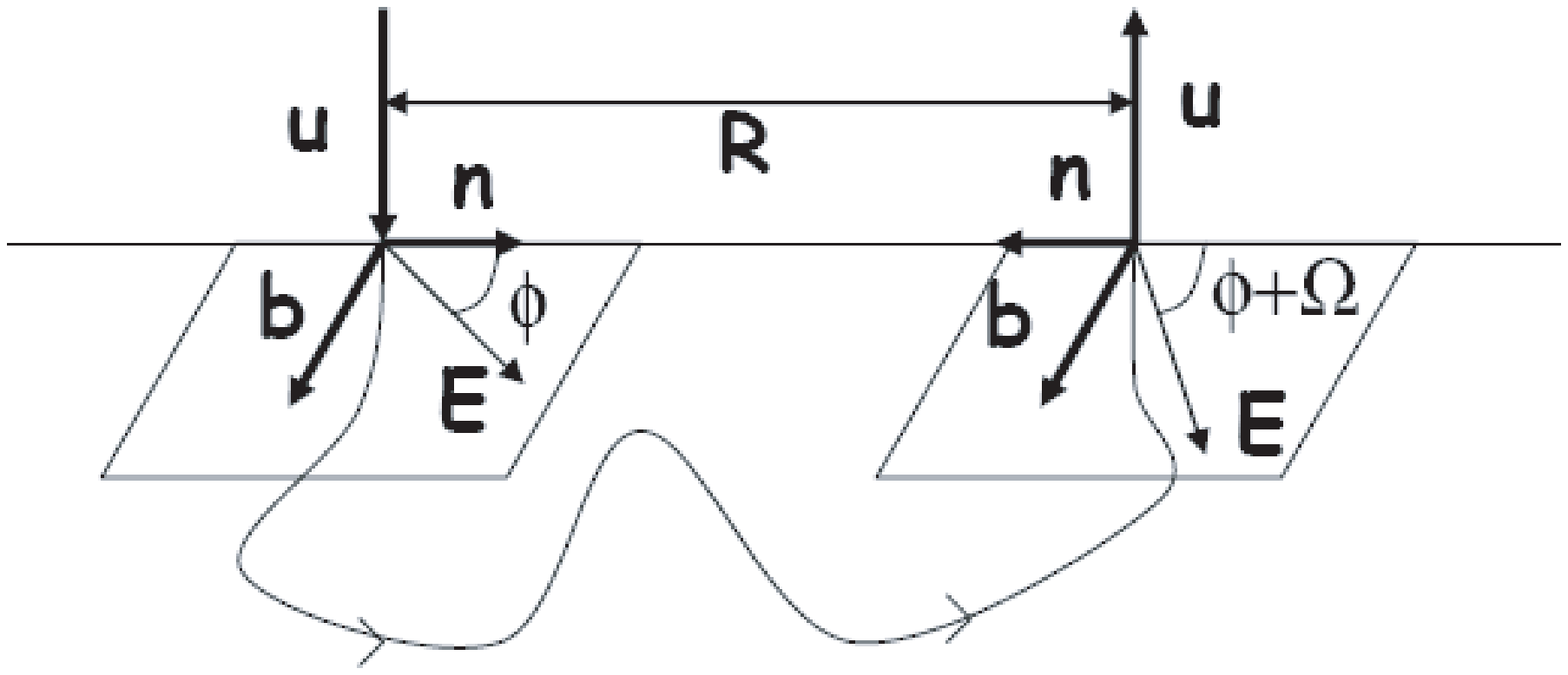}}}
\par}
\caption{Representation of a typical path in a semi-infinite
random medium in backscattering. The Fr\'enet frame consists of
the tangent $\bf u$, the normal $\bf n$, and the binormal $\bf b$
vectors. $R$ denotes the distance between end points, $\phi$ is
the initial angle between the polarization vector $\bf E$ and the
normal $\bf n$, and $\Omega$ is the geometric phase.}
\end{figure}

\newpage
\begin{figure} \label{fig:dist-s}
{\par\centering
{\rotatebox{0}{\includegraphics[scale=0.2]{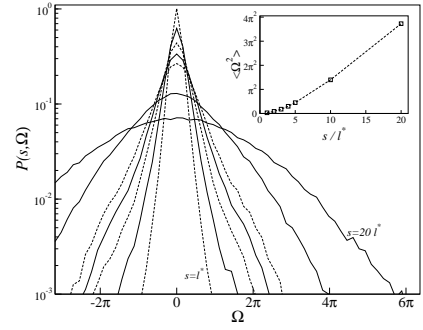}}}
\par}
\caption{Distribution of the geometric phase $\Omega$ for
different values of the path length $s$ and in the inset variance
of the distribution as function of $s/\ell^*$.}
\end{figure}

\newpage
\begin{figure} \label{fig:dist-c}
{\par\centering
{\rotatebox{-90}{\includegraphics[scale=0.5]{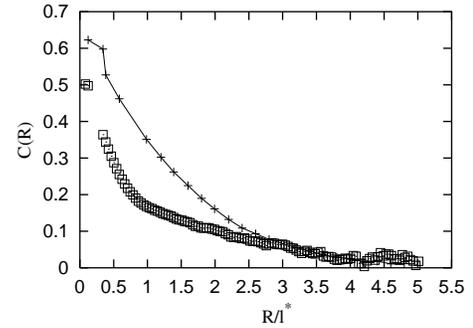}}}
\par}
\caption{Contrast as function of $R$: the crosses have been
obtained from Monte-Carlo simulations using Eq.~\ref{CofR} and the
squares are experimental values, obtained from an analysis of the
Stokes parameter $Q$.}
\end{figure}

\end{document}